\documentstyle[12pt,psfig,epsf]{article}
\newcommand{\beq}[1]{\begin{equation}\label{#1}}
\newcommand{\eeq}{\end{equation}}
\newcommand{\beqar}[1]{\begin{eqnarray}\label{#1}}
\newcommand{\eeqar}{\end{eqnarray}}
\newcommand{\ep}{\varepsilon}

\newcommand{\La}{\Lambda}

\begin{document}
\vspace*{-2cm}
\hfill UFTP preprint 414/1996
\vspace{4cm}
\begin{center}
{\LARGE \bf IR-Renormalon Contribution to the\\[2mm]
Polarized Structure Function $g_1$}
\vspace{1cm}

M.~Meyer-Hermann$^\dagger$, 
M.~Maul$^\dagger$,
L.~Mankiewicz$^{\dagger\dagger}$, 
E.~Stein$^\dagger$,
and A.~Sch\"afer$^\dagger$
\vspace{1cm}

$^\dagger$Institut f\"ur Theoretische Physik, J.~W.~Goethe
Universit\"at Frankfurt,
\\ 
Postfach~11~19~32, D-60054~Frankfurt am Main, Germany
\\ \vspace{2em}
$^{\dagger\dagger}$Institut f\"ur Theoretische Physik, TU-M\"unchen, 
D-85747 Garching, Germany\footnote{On leave of absence 
from N.Copernicus Astronomical Center, Bartycka 18, PL--00--716 Warsaw,
Poland}
\end{center}

\vspace{2cm}
\noindent{\bf Abstract:}
An estimation of the higher twist contribution to
the polarized nonsinglet structure function $g_1$ is given. 
Its first moment is generally assumed to be small but this could be due
to a cancellation of its contribution in different Bjorken-$x$ ranges.
We estimate the magnitude of $g_1^{tw4}$ as a function of 
$x$ in the framework of the renormalon method. 
The calculated
higher-twist-correction to $g_1$ turn out to be in fact small for all
$x$ values. The correction to the determination of the nonsinglet
matrix element
$d_{NS}^{(2)}$  from data for transversely nucleon polarization due
to the calculated higher twist effects is of order $10\%$.
\vspace*{\fill}
\eject
\newpage

Recently the E143 collaboration measured the transverse polarized
structure function for proton and neutron \cite{E143} and obtained
a first approximate result for the nonsinglet twist-3 matrix element
$d_{NS}^{(2)}$. This analysis relies on the assumption that the
higher twist contributions to $g_1^{p-n}$ are negligible.
This experimental result is in strong disagreement with predictions from
lattice calculations \cite{Goe96}, and in marginal agreement
with predictions from QCD sum rule calculations \cite{Ste95a}
such that possible sources of error have to be looked into very
carefully. One such uncertainty in the experimental determination of
$d_{NS}^{(2)}$ is the size of higher twist contributions to
$g_1^{p-n}$. Theoretically they could be large even if they
give only a small contribution to the first moment of
$g_1^{p-n}$, for example if the latter is
due to a cancellation of twist-4 contributions
from different Bjorken-$x$ regions with different signs.

To know the size of the higher twist contributions is  also very
important for recent attempts to obtain $\Delta G(x)$ from the
$Q^2$-dependence of $g_1(x,Q^2)$ \cite{Bal95a,Lic95}.
We present a renormalon estimate showing that there are in fact
cancelations for the first moment of $g_1^{tw4}$. The higher
twist contributions to $g_1(x,Q^2)$ are nevertheless small, such
that the mentioned analysis of experimental data is valid.

Neglecting higher twist contributions
the moments of the nonsinglet polarized structure function $g_1(x,Q^2)$
can, in the framework of operator product expansion, be identified
with the twist-2 nonsinglet operator matrix elements $A_{g_1,n}$
(we abbreviate $g_1(x,Q^2) := g_1^p(x,Q^2) - g_1^n(x,Q^2)$):
\beqar{Mg1}
M_{g_1,n}(Q^2)
 &=& \int_0^1 dx\;x^{n-1} g_1(x,Q^2) \nonumber\\
 &=& C_{g_1,n}\left(\frac{Q^2}{\mu^2},a_s\right) A_{g_1,n}(\mu^2)
\eeqar
Here we introduced the factorization scale $\mu$ and
the Wilson-coefficient $C_n$.
$a_s$ is the strong coupling, defined by
$a_s := \frac{g^2}{(4\pi)^2}$.
Higher twist operator matrix elements are not written.

To estimate higher twist contributions to eq.\,(\ref{Mg1})
we use the renormalon-method \cite{Mue92}.
The forward Compton-scattering-amplitude is calculated in the Borel-plane
with an effective gluon propagator, taking only one gluon exchange 
into account.
In the Landau gauge 
the effective gluon propagator resums arbitrary many quark
loop insertions in the gluon line, which is an exact procedure
in QED.
The gluon propagator reads in the Borel representation \cite{Ben94}:
\beq{gluon}
{\cal B}_{1/a_s}[a_s D^{ab}_{\mu\nu}(k)](u)
\;=\;
\delta^{ab}\frac{ g_{\mu\nu} - \frac{k_\mu k_\nu}{k^2}}{k^2}
\left(\frac{\mu^2 e^{-C}}{-k^2}\right)^{\beta_0 u} \; ,
\eeq
$C$ corrects for
the renormalization scheme dependence ($C=-\frac{5}{3}$ for
$\overline{MS}$-scheme) and
$u$ is the Borel transformation parameter. In QCD
the restriction to one gluon
exchange is an exact procedure in the large $N_F$-limit
\cite{Bro93,Kat93,Lov95}, where $N_F$ is
the number of quark-flavors. 
The next-to-leading $N_F$-order terms are
approximated by naive-nonabelianization (NNA) 
\cite{Ben94,Bro95,Ben95,Bal95}, 
which means 
to replace the one loop beta-function of QED 
by the QCD beta-function $\beta_0 = 11 - \frac{2}{3} N_F$ 
and corresponds to
the replacement $N_F \rightarrow N_F - \frac{33}{2}$.
The quality of this
approximation has to be checked by comparing the
NNA-perturbative-coefficients with the known
exact ones.

The IR-renormalon-poles lead to an factorial growth of the
perturbative series, 
which has to be interpreted as an asymptotic expansion.
In agreement with the general theory of asymptotic
series it has to be truncated after the minimal term, which
determines the best accuracy
which can be achieved using perturbative expansion. 
As its $Q^2$ dependence is
power-like, it has been suggested to use it as an estimate of
higher-twist contributions. 
Despite 
the fact that the conceptual basis for this approach is controversial
\cite{Dok95}, the procedure has given previously reasonable
estimates \cite{Bra95,Dok95b,Ste96} and is much simpler than
genuine higher twist estimates due to lattice or QCD sum rules 
calculations.

Finally, let us mention that there are two popular methods of computing the
renormalon ambiguity in the QCD perturbative series. 
Apart from the
Borel-transformed propagator (\ref{gluon}) one can use
a gluon propagator with a small gluon mass $\lambda^2$.
The coefficient of the
$n$-th renormalon pole is the same as the coefficient of the
$\lambda^{2n} ln(\lambda^2)$-term in the small gluon-mass approach.
Despite seemingly different
physical meaning, both methods are exactly 
equivalent \cite{Bal95,Ben94a}.

The paper is organized as follows: first the Wilson-coefficient for all
moments of $g_1$ are calculated in the NNA approximation using the
Borel-transformed gluon-propagator. 
Then the perturbative coefficients
are calculated in the NNA approximation and are compared to known
exact values. 
Next the IR renormalon ambiguity is calculated and used 
for an estimate of the twist-4 corrections to the polarized structure
function. As the calculation is done for all moments, the twist-4
corrections are given as a function of Bjorken-$x$. Finally we give
a phenomenological analysis of the twist-4 contribution.


The forward Compton scattering amplitude corresponding to the polarized
structure function $g_1$ reads after an expansion in $\omega=1/x$:
\beq{Tmunu}
T_{\mu\nu}^{g_1}
\;=\;
2i\ep_{\mu\nu\lambda\rho} \frac{q^\lambda s^\rho}{p \cdot q}
\sum_{\stackrel{n=1}{\rm odd}}^\infty
C_{g_1,n}\left(\frac{Q^2}{\mu^2},a_s\right) A_{g_1,n}(\mu^2)
\,\omega^n
\eeq
We have computed the one-gluon-exchange diagrams contributing to
deep inelastic lepton-nucleon scattering using the 
effective gluon propagator (\ref{gluon}). 

The result in the Borel plane is expanded in $\omega$ and compared to
eq.\,(\ref{Tmunu}) to read off the Borel transformed Wilson 
coefficient. Note
that the Borel transformation of eq.\,(\ref{Tmunu}) 
is given by a similar expression with the Borel transformed 
Wilson coefficient on the
right hand side. The resulting antisymmetric part is 
(we use the notation of Muta \cite{Mut87} and $s=\beta_0 u$):
\beqar{wilson}
{\cal B}_{\frac{1}{a_s}}
\!\left[C_{g_1,n}\left(\frac{Q^2}{\mu^2},a_s\right)\right]\!(u)
&\!\!\!=\!\!\!&
-2C_F\left(\frac{\mu^2 e^{-C}}{Q^2}\right)^s \nonumber\\
&&
\Bigg\{
  \frac{1}{s}
  \bigg[ \frac{\Gamma(s+n)}{\Gamma(1+s) \Gamma(n)}
         \left( \frac{1}{(s+n)(1+s+n)} + \frac{1}{1+s+n} + \frac{1}{2}\right)
\nonumber\\
&&\hspace{6mm}-
         \sum_{k=1}^n
         \frac{\Gamma(s+k)}{\Gamma(1+s) \Gamma(k)}
         \left( \frac{1}{s+k} + \frac{1}{1+s+k} \right)
  \bigg]
\nonumber\\
&&+
  \frac{2}{1-s}
  \bigg[ \frac{\Gamma(s+n)}{\Gamma(1+s) \Gamma(n)}
         \left( \frac{1}{(s+n)(1+s+n)} + \frac{1}{1+s+n} + \frac{1}{2}\right)
\nonumber\\
&&\hspace{13mm}-
         \sum_{k=1}^n
         \frac{\Gamma(s+k)}{\Gamma(1+s) \Gamma(k)}
         \frac{1}{1+s+k}
  \bigg]
\nonumber\\
&&-
  \frac{1}{2-s}
  \bigg[ \frac{\Gamma(s+n)}{\Gamma(1+s) \Gamma(n)}
         \left(\frac{1}{1+s+n} + \frac{1}{2}\right)
\nonumber\\
&&\hspace{13mm}+
         \sum_{k=1}^n
         \frac{\Gamma(s+k)}{\Gamma(1+s) \Gamma(k)}
         \left( \frac{1}{s+k} - \frac{1}{1+s+k} \right)
  \bigg]
\Bigg\}
\eeqar
For the first moment this result coincides with the one obtained
by \cite{Kat93}.
There are IR-renormalons appearing at $s=1,2$ and UV-renormalons depending
on the considered moment of $g_1$. The fact that there are only $2$ 
IR-renormalons appearing is general for the case of deep inelastic
lepton-nucleon scattering, but not imperative for other processes.
The $1/s$-term has still to be renormalized by a corresponding
counterterm, which, to lowest order, is given by
the coefficient in front of the $1/s$ singularity:
\beq{anomalous}
C_{1/s}
\;=\;
-2C_F\left(\frac{1}{n(1+n)} - \frac{1}{2} -2\sum_{k=2}^n \frac{1}{k}
      \right)
\;=\; -2 \gamma_n^\Delta
\eeq
where $\gamma_n^\Delta$ is the anomalous dimension \cite{Cra83}
corresponding to the parton to parton branching
with helicity transfer. 
This expression vanishes for $n=1$, so that there exists
no $1/s$ singularity in the case of the Bjorken sum rule.
The factor $2$ is due to the fact that we defined
$a_s=\alpha_s/(4\pi)$ instead of $\alpha_s/(2\pi)$. The relative
sign can be understood as follows: The coefficient of
$1/s$ has to be identified with the coefficient of 
$\ln(\mu^2)$ \cite{Ben95}, 
while the anomalous dimension appears in front of
$\ln(Q^2/\mu^2)=\ln(Q^2)-\ln(\mu^2)$. 
This gives rise to the minus sign.


The Bjorken sum rule has been determined perturbatively up to high orders
in $a_s$ \cite{Lar91}. Writing the result in form of an expansion
in $N_F$ one gets:
\beqar{Bjorken}
\int_0^1 dx\,g_1(x,Q^2)
&=&
\frac{1}{6}\left|\frac{g_A}{g_V}\right|
\bigg\{
1-4a_s - \left(\frac{220}{3} - \frac{16}{3}N_F\right) a_s^2
\nonumber\\
&&\hspace{1.4cm}
- \left(2652.15 - 486.87N_F + \frac{920}{81}N_F^2\right) a_s^3
+ {\cal O}(a_s^4) \bigg\}
\nonumber\\
&=&
\frac{1}{6}\left|\frac{g_A}{g_V}\right|
\left\{
1-4a_s - 57.33\; a_s^2
- 1302.76\; a_s^3
+ {\cal O}(a_s^4) \right\}
\eeqar
where the Riemann zeta functions are given in decimal form and in the
last line $N_F=3$ was inserted.

The perturbative corrections to the Bjorken sum rule can be reconstructed
from the Borel transformed Wilson coefficient $C_{g_1,1}$ 
in eq.\,(\ref{wilson})
in the NNA approximation by successive differentiation 
with respect to $s$.
Let $m$ be the order of the perturbative correction and $B_n^{(m)}$ the
corresponding coefficient. Then 
\beq{perturbativ}
B_{g_1,1}^{(m+1)} \;=\; 
\left.
\beta_0^m \frac{d^m}{ds^m}
 {\cal B}_{\frac{1}{a_s}}
\!\left[C_{g_1,1}\left(Q^2=\mu^2,a_s\right)\right](s)
\right|_{s=0}
\eeq
This equation remains valid for higher moments if the renormalized
Wilson coefficient is inserted on the right hand side.
In the $\overline{MS}$ scheme we get for the NNA approximated perturbative
coefficients:
\beqar{NNABjorken}
\int_0^1 dx\,g_1(x,Q^2)
&=&
\frac{1}{6}\left|\frac{g_A}{g_V}\right|
\bigg\{
1-4a_s - \left(\frac{264}{3} - \frac{16}{3}N_F\right) a_s^2
\nonumber\\
&&\hspace{1.4cm}
- \left(3092.22 - 374.81N_F + \frac{920}{81}N_F^2\right) a_s^3
+ {\cal O}(a_s^4) \bigg\}
\nonumber\\
&=&
\frac{1}{6}\left|\frac{g_A}{g_V}\right|
\left\{
1-4a_s - 72\; a_s^2 - 2070\; a_s^3
+ {\cal O}(a_s^4) \right\}
\eeqar
where we inserted $N_F=3$ in the last line.
So the contributions from leading order in $N_F$ are reproduced exactly, while
the nonleading terms are found to be approximated with an relative error of
about $30\%$ (see also \cite{Lov95}).


As already mentioned the terms in perturbative 
series in QCD have the general
behaviour to reach a minimum at some order $m_0$ 
of the coupling and to
diverge for orders $m>m_0$. 
So the series has to be truncated at the order
$m_0$ to get a meaningful result.
\beq{truncate}
C_{g_1,n}\left(Q^2=\mu^2,a_s\right)
\;=\;
\sum_{k=0}^{m_0} B_{g_1,n}^{(k)} a_s^k
+
C_{g_1,n}^{(1)} \frac{\Lambda_C^2 e^{-C}}{Q^2}
+
C_{g_1,n}^{(2)} \frac{\Lambda_C^4 e^{-2C}}{Q^4}
\eeq
The ambiguity of this procedure is just
of the order of the smallest term and has the form of a power suppressed
contribution. In the Borel plane this uncertainty 
is related to an IR-renormalon
pole, which prevents the unambiguous reconstruction of the resummed
series. This ambiguity appears as the
imaginary part of the Laplace integral:
\beq{inverse-borel}
\sum_k
C_{g_1,n}^{(k)} \left(\frac{\Lambda_C^2 e^{-C}}{Q^2}\right)^k
\;=\;
\frac{1}{\pi \beta_0}
{\rm Im} \int_0^\infty ds\;e^{-s/(\beta_0 a_s)}
{\cal B}_{\frac{1}{a_s}}
\!\left[C_{g_1,n}\left(\frac{Q^2}{\mu^2},a_s\right)\right](s)
\eeq
Using the Borel transformed Wilson coefficient (\ref{wilson})
we get for the leading IR-renormalon ambiguity
\beq{s=1pol}
C_{g_1,n}^{(1)}
\;=\;
\pm \frac{4 C_F}{\beta_0}
\left\{\frac{n(3+n)}{2(1+n)} - \sum_{k=1}^n \frac{k}{2+k}
\right\}
\eeq
The formally undetermined sign is due to the two possibilities to choose the
integration path in eq.\,(\ref{inverse-borel})
avoiding the renormalon singularity. 
We checked in the case of the Bjorken sum rule, that this result is
reproduced up to logarithms by the explicit calculation
of the minimal term in the asymptotic perturbative series. 
It is tempting to assume that the resulting sign of the higher-twist
contribution is identical to that of the minimal term in the
perturbative series generated by the IR-renormalon.
In earlier applications this procedure lead to results, which agree
with experimental data \cite{Ste96}. 
Accordingly we suppose that the sign in eq.\,(\ref{s=1pol})
should be negative, 
but strictly speaking it remains unknown.
%

In an analogous way the nonleading IR-renormalon contribution is calculated:
\beq{s=2pol}
C_{g_1,n}^{(2)}
\;=\;
\pm \frac{2C_F}{\beta_0}
\left\{\frac{n(1+n)}{4} + \frac{1}{3} + \sum_{k=2}^n \frac{k}{2+k}
\right\}
\eeq
Note that it is not clear at this point whether this contribution is more
important than the $a_s$-correction to the leading power correction 
or not.

To use this result for an estimate of power corrections
to $g_1$ as a function of Bjorken-$x$, eq.\,(\ref{s=1pol}) and
(\ref{s=2pol}) are transformed back to momentum space
\beq{C1(x)}
C_{g_1}^{'(1)} (x)
\;=\;
\pm \frac{2C_F}{\beta_0}
\left\{
4 + 2x + 4x^2 - \frac{4}{(1-x)_+} + \delta'(x-1) - 5\delta(x-1)
\right\}
\eeq
and
\beq{C2(x)}
C_{g_1}^{'(2)} (x)
\;=\;
\pm \frac{C_F}{\beta_0}
\left\{
-4 - 4x - 4x^2 + \frac{4}{(1-x)_+} 
+ 9\delta(x-1) - 4\delta'(x-1)  + \frac{1}{2} \delta''(x-1)
\right\}
\eeq
with
\beq{momC1}
C_{g_1,n}^{(i)}
\;=\;
\int_0^1 dx\; x^{n-1}
C_{g_1}^{'(i)} (x)
\eeq
and $1/(1-x)_+$ is the singularity subtracted contribution to the integral
(\ref{momC1}). 

While writing up this contribution
we learned of an independent calculation \cite{Das96}
for this quantity using the equivalent small gluon mass approach.
Both calculations agree completely.


\begin{figure}[ht]
\vspace{1cm}
\centerline{\psfig{figure=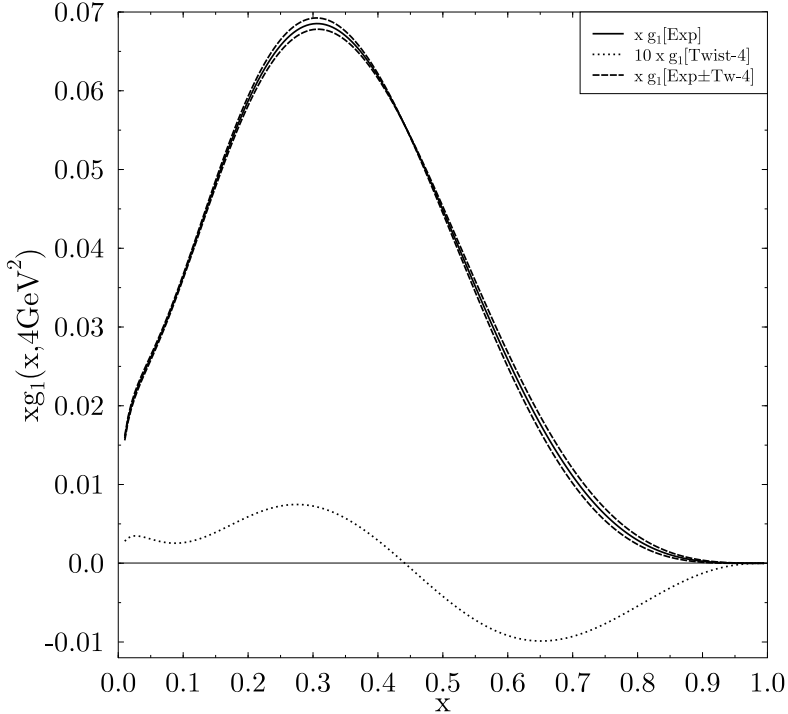,width=15cm}}
\caption[]{\sf
The experimental fit for $g_1(x,4\,{\rm GeV}^2)$ (full line)
\cite{Geh95}. The estimate
for the twist-4 correction is given explicitly multiplied by a
factor of $10$ (dotted line), showing a change in sign at 
$x\approx 0.43$. The correction was
added and subtracted from the
experimental values (dashed lines), so that the theoretical prediction
of the twist-2 part of $g_1$ has to lie within this region
($\La_{\overline{MS}}=200~{\rm MeV}$, 
$Q^2=4~{\rm GeV}$ and $N_f = 4$).}
\label{fig1}
\end{figure}

\begin{figure}[ht]
\vspace{1cm}
\centerline{\psfig{figure=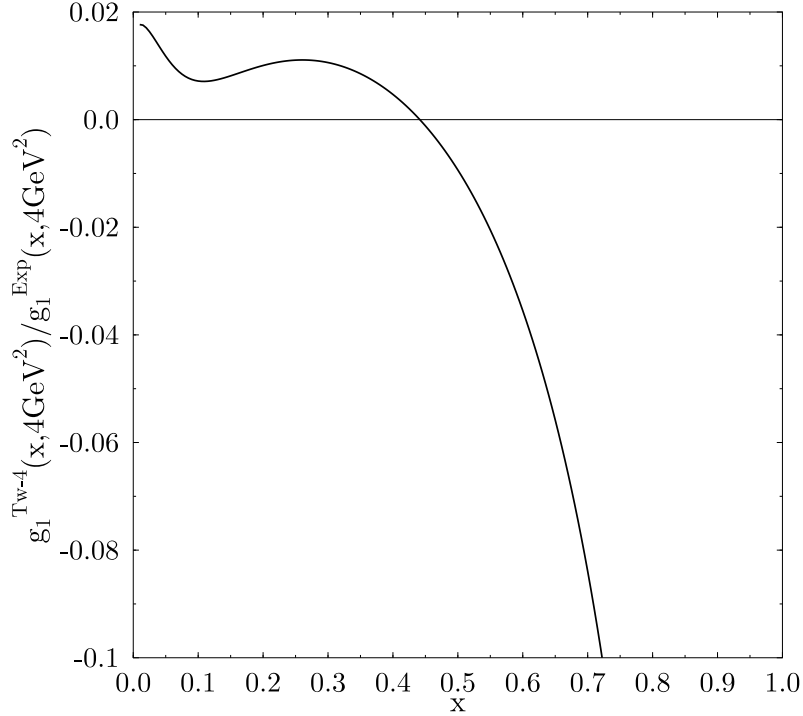,width=15cm}}
\caption[]{\sf
The relative magnitude of the predicted twist-4 contribution to 
$g_1$ with respect to the experimental fit.
$\La_{\overline{MS}}=200~{\rm MeV}$, $Q^2=4~{\rm GeV}$ and $N_f = 4$.}
\label{fig2}
\end{figure}

The ratio of the moments $M_{g_1,n}^{tw4}$
of the twist-4 part of $g_1$ and $M_{g_1,n}^{exp}$ of the experimental
values for $g_1$ (containing all twist) reads
\beqar{ratio}
\frac{M_{g_1,n}^{tw4}(Q^2)}{M_{g_1,n}^{exp}(Q^2)}
&=&
\frac{C_{g_1,n}^{(1)} \frac{\Lambda_C^2 e^{-C}}{Q^2}
      + {\cal O}\left(\frac{a_s}{Q^2}, \frac{1}{Q^4}\right)}
     {\sum_{k=0}^{m_0} B_{g_1,n}^{(k)} a_s^k +
      C_{g_1,n}^{(1)} \frac{\Lambda_C^2 e^{-C}}{Q^2}
      + {\cal O}\left(\frac{a_s}{Q^2}, \frac{1}{Q^4}\right)}
\nonumber\\
&=&
C_{g_1,n}^{(1)} \frac{\Lambda_C^2 e^{-C}}{Q^2}
+ {\cal O}\left(\frac{a_s}{Q^2}, \frac{1}{Q^4}\right) \, ,
\eeqar
where we used $B_{g_1,n}^{(0)}=1$.
In this way $g_1^{tw4}(x,Q^2)$ is expressed by a convolution of
$g_1^{exp}(x,Q^2)$:
\beqar{convolution}
g_1^{tw4}(x,Q^2)
&\!\!\!=\!\!&
\frac{\Lambda_C^2 e^{-C}}{Q^2}
\int_x^1 \frac{dy}{y}\, g_1^{exp}\left(\frac{x}{y}\right)
                       \, C_{g_1}^{'(1)} (y)
\nonumber\\
&\!\!\!=\!\!&
\frac{2C_F}{\beta_0} \frac{\Lambda_C^2 e^{-C}}{Q^2}
\bigg\{
\int_x^1 \frac{dy}{y}\, g_1^{exp}\left(\frac{x}{y}\right)
                          (4+2y+4y^2)
-4 \int_x^1 \frac{dy}{y}\, g_1^{exp}\left(\frac{x}{y}\right)
                              \frac{1}{(1-y)_+}
\nonumber\\
&&\hspace{2.2cm}
+ x \frac{d}{dx}g_1^{exp}(x) - 4 g_1^{exp}(x)
\bigg\}
\eeqar
where $C_{g_1}^{'(1)}(y)$ eq.\,(\ref{C1(x)}) was inserted.

Using a fit to the available experimental data for $g_1$ \cite{Geh95},
this expression indeed leads to very small twist-4 corrections to $g_1(x)$
(see fig.\,\ref{fig1}, \ref{fig2}). This result is not significant for
the small $x$ region, where the large-$N_F$ limit is no longer justified.

From our general result for $g_1^{tw4}(x,Q^2)$ eq.\,(\ref{convolution})
we can
also reproduce the renormalon prediction for the nonsinglet
twist-4 matrix element \cite{Bra95}
\beq{f2tw4}
\int_0^1 dx\;g_1^{tw4}(x,Q^2)
\;=\; 
\pm \frac{4C_F}{3\beta_0} \frac{\Lambda_C^2 e^{-C}}{Q^2}
\int_0^1 dx\;g_1^{exp}(x,Q^2)
\;\approx\;
\pm \frac{0.017\,GeV^2}{Q^2}
\eeq
in good agreement with our QCD sum rule result of 
$-\frac{0.0094\,GeV^2}{Q^2}$ \cite{Ste95b}. 
Another renormalon estimate based on Pad\'e approximants
\cite{Ell96} gave a value of $\frac{\pm (0.04 \pm 0.016)\,GeV^2}{Q^2}$.
In the same way as we did in \cite{Ste96}
we checked the cancellation of the above IR-renormalon
contribution by the UV-renormalon contribution to
the genuine twist-4 correction. Note that in addition
to the twist-4 power correction to the Bjorken sum rule
there exist twist-2 and twist-3 matrix elements, which are also
power suppressed by $1/Q^2$ \cite{Shu82}. 

Finally we give an estimate for the error made in the determination
of the nonsinglet twist-3 matrix element 
$d^{(2)}$ due to higher-twist corrections.
The third moment of the nonsinglet structure function 
$g_2(x,Q^2) := g_2^p(x,Q^2) - g_2^n(x,Q^2)$ is
given by
\beq{mom3g2}
\int_0^1 dx\,x^2 g_2(x,Q^2)
\;=\; - \frac{1}{3}a^{(2)} + \frac{1}{3} d^{(2)}
      + {\cal O}\left(\frac{M^2}{Q^2}\right)
\eeq
where $a^{(2)}$ is a pure twist-2 nonsinglet operator matrix element. 
So the twist-3 nonsinglet matrix element $d^{(2)}$
can be obtained by measuring $g_2(x,Q^2)$ and $a^{(2)}$, which in 
turn is determined by the third moment of $g_1(x,Q^2)$:
\beq{ellismom3g1}
\int_0^1 dx\,x^2 g_1(x,Q^2)
\;=\;  \frac{1}{2}a^{(2)}
      + {\cal O}\left(\frac{M^2}{Q^2}\right)
\eeq
If one uses $a^{(2)}$ from the total measured $g_1(x,Q^2)$,
higher twist corrections to $g_1(x,Q^2)$ are
not distinguished from the twist-2 part.
This leads to an error in
the determination of $d^{(2)}$ of the order of the higher-twist
correction to $g_1(x,Q^2)$ \cite{E143}. 
Using our estimate for the nonsinglet
twist-4 contribution to $g_1(x,Q^2)$, we obtain the following uncertainty
for $a^{(2)}$
\beq{errora2}
2\int_0^1 dx\;x^2 g_1^{tw4}(x,Q^2) \approx 0.0003 
\sim \Delta(a^{(2)})
\eeq
The main contribution to the third moment of $g_2(x,Q^2)$ comes from
$a^{(2)}$, so that the above uncertainty leads to an error for
$d^{(2)}$ of about $10\%$, which is substantially smaller than 
the present experimental
error bars. We conclude that the
mentioned procedures to determine $d^{(2)}$ and $\Delta G(x)$
from experimental data are not invalidated by higher-twist effect.

\vspace*{5mm}
{\bf Acknowledgements.} This work has been  
supported by  BMBF, DFG (G.Hess Programm) and KBN
grant 2~P03B~065~10. A.S.~thanks also the
MPI f\"ur Kernphysik in Heidelberg for support.

\vfill
\eject

\end{document}